\pdfoutput=1
\documentclass[longauth]{aa}  
\usepackage{natbib}
\bibpunct{(}{)}{;}{a}{}{,}

\usepackage{graphicx}
\usepackage{txfonts}
\begin{document}

   \title{MAGIC reveals a complex morphology within the unidentified gamma-ray source HESS J1857+026}
  
\author{
J.~Aleksi\'c\inst{1} \and
S.~Ansoldi\inst{2} \and
L.~A.~Antonelli\inst{3} \and
P.~Antoranz\inst{4} \and
A.~Babic\inst{5} \and
P.~Bangale\inst{6} \and
U.~Barres de Almeida\inst{6} \and
J.~A.~Barrio\inst{7} \and
J.~Becerra Gonz\'alez\inst{8,}\inst{25} \and
W.~Bednarek\inst{9}
% ...
}
\institute { IFAE, Campus UAB, E-08193 Bellaterra, Spain
\and Universit\`a di Udine, and INFN Trieste, I-33100 Udine, Italy
\and INAF National Institute for Astrophysics, I-00136 Rome, Italy
\and Universit\`a  di Siena, and INFN Pisa, I-53100 Siena, Italy
\and Croatian MAGIC Consortium, Rudjer Boskovic Institute, University of Rijeka and University of Split, HR-10000 Zagreb, Croatia
\and Max-Planck-Institut f\"ur Physik, D-80805 M\"unchen, Germany
\and Universidad Complutense, E-28040 Madrid, Spain
\and Inst. de Astrof\'isica de Canarias, E-38200 La Laguna, Tenerife, Spain
\and University of \L\'od\'z, PL-90236 Lodz, Poland
\and Deutsches Elektronen-Synchrotron (DESY), D-15738 Zeuthen, Germany
\and ETH Zurich, CH-8093 Zurich, Switzerland
\and Universit\"at W\"urzburg, D-97074 W\"urzburg, Germany
\and Centro de Investigaciones Energ\'eticas, Medioambientales y Tecnol\'ogicas, E-28040 Madrid, Spain
\and Technische Universit\"at Dortmund, D-44221 Dortmund, Germany
\and Inst. de Astrof\'isica de Andaluc\'ia (CSIC), E-18080 Granada, Spain
\and Universit\`a di Padova and INFN, I-35131 Padova, Italy
\and Universit\`a dell'Insubria, Como, I-22100 Como, Italy
\and Unitat de F\'isica de les Radiacions, Departament de F\'isica, and CERES-IEEC, Universitat Aut\`onoma de Barcelona, E-08193 Bellaterra, Spain
\and Institut de Ci\`encies de l'Espai (IEEC-CSIC), E-08193 Bellaterra, Spain
\and Japanese MAGIC Consortium, Division of Physics and Astronomy, Kyoto University, Japan
\and Finnish MAGIC Consortium, Tuorla Observatory, University of Turku and Department of Physics, University of Oulu, Finland
\and Inst. for Nucl. Research and Nucl. Energy, BG-1784 Sofia, Bulgaria
\and Universitat de Barcelona, ICC, IEEC-UB, E-08028 Barcelona, Spain
\and Universit\`a di Pisa, and INFN Pisa, I-56126 Pisa, Italy
\and now at: NASA Goddard Space Flight Center, Greenbelt, MD 20771, USA and Department of Physics and Department of Astronomy, University of Maryland, College Park, MD 20742, USA
% ...
}

   \date{Received: 27 January 2014}
   
   \offprints{V.~Stamatescu (vstamatescu@ifae.es), J.~Krause (julkrau@googlemail.com) and S.~Klepser (klepser@ifae.es)}

   
  \abstract
  {}
   {\mbox{HESS J1857+026} is an extended TeV gamma-ray source that was discovered by H.E.S.S. 
    as part of its Galactic plane survey. Given its broadband spectral energy distribution
    and its spatial coincidence with the young energetic pulsar \mbox{PSR J1856+0245},
    the source has been put forward as a pulsar wind nebula (PWN) candidate.
    MAGIC has performed follow-up observations aimed at mapping
    the source down to energies approaching 100 GeV in order to better understand its complex morphology.}
   {\mbox{HESS J1857+026} was observed by MAGIC in 2010, yielding 29 hours of good quality stereoscopic data
     that allowed us to map the source region in two separate ranges of energy.}
   { 
     We detected very-high-energy gamma-ray emission from \mbox{HESS J1857+026} with a significance of $12 \sigma$ above $150$ GeV.
     The differential energy spectrum between $100$ GeV and $13$ TeV is well described by a power law function $dN/dE = N_0(E/1\textrm{TeV})^{-\Gamma}$ with 
     $N_0 = (5.37 \pm0.44_{stat} \pm1.5_{sys}) \times 10^{-12} (\textrm{TeV}^{-1} \textrm{cm}^{-2}$ $\textrm{ s}^{-1})$ and $\Gamma = 2.16\pm0.07_{stat} \pm0.15_{sys}$,
     which bridges the gap between the GeV emission measured by \textit{Fermi}-LAT and the multi-TeV emission measured by H.E.S.S..
     In addition, we present a detailed analysis of the energy-dependent morphology of this region.
     We couple these results with archival multi-wavelength data and outline evidence in favor of a two-source scenario,
     whereby one source is associated with a PWN, while the other could be linked with a molecular cloud complex containing an H{\sc ii} region and a possible gas cavity.}
   {}
   
   
   \authorrunning{MAGIC Collaboration}
   \titlerunning{}
   \maketitle
%

\section{Introduction}

   One of the long standing goals of very-high-energy (VHE) gamma-ray astronomy
   is to trace the particle populations responsible for producing TeV photons
   and, in so doing, to search for the sources of cosmic rays in our Galaxy.
   VHE gamma rays may be produced through hadronic interactions
   via the $\pi^0$ decay channel and by energetic electrons through inverse compton (IC) scattering
   on soft photon fields or via nonthermal Bremsstralung.
   These parent particles are thought to be energized by shock acceleration,
   which very likely happens in astrophysical objects
   such as supernova remnants (SNRs), pulsar wind nebulae (PWNe) and compact binaries.
   More recently, some Galactic TeV gamma-ray sources have been linked
   to regions of massive star formation,
   although these associations are complicated
   by the expected presence of additional objects of the aforementioned types.
   Finally, nearly one third of all Galactic sources, many of which were discovered
   through the H.E.S.S. galactic plane survey \citep{Carrigan2013} remain unidentified
   and could provide new insight into these extreme astrophysical environments.
   
   \mbox{HESS J1857+026} was discovered as a source without clear associations at other wavelengths
   during the H.E.S.S. survey of the inner Galaxy \citep{Aharonian2008}.
   The source had an intrinsic extension of
   $(0.11 \pm 0.08_{stat})^{\circ} \times (0.08 \pm 0.03_{stat})^{\circ}$,
   with an inclination (measured counter-clockwise with
   respect to the RA-axis) of $(-3 \pm 49_{stat})^{\circ}$.
   A significant tail-like structure toward the north was also seen,
   which hinted at either a larger overall extension or a weaker but distinct northern source.
   The spectrum of \mbox{HESS J1857+026}, which covered the range of $0.8-45$ TeV,
   was well described by a single power law with a spectral index of $2.39 \pm 0.08_{stat}$.
     
   Subsequent searches at radio wavelengths have led to the discovery
   of a nearby energetic pulsar, \mbox{PSR J1856+0245},
   with a spin period of 81 ms and characteristic spin-down age of 21 kyr \citep{Hessels2009}. 
   Given a pulsar dispersion measure (DM) of $622 \pm2$ cm$^{-3}$pc,
   the distance was estimated to be $\sim9$ kpc
   based on the NE2001 model for the Galactic distribution of free electrons \citep{CordesLazio2002}.
   The uncertainty on this distance is not well determined and the value may
   vary by factors of $2-3$ \citep{Hessels2009},
   which in turn leads to large uncertainties on the estimated size
   and energetics of the candidate PWN.

   \mbox{HESS J1857+026} was also detected at GeV energies \citep{NeronovSemikoz2010,Paneque2011,Rousseau2012},
   although no significant pulsation from the pulsar was found in 3 years of \textit{Fermi}-LAT data.
   \citet{Rousseau2012} extracted a spectrum assuming a point source,
   and fit both leptonic and hadronic models to a broadband spectral energy distribution (SED)
   that used the original H.E.S.S. results and preliminary results from MAGIC \citep{Klepser2011}.
   Their multi-wavelength SED also included an upper limit on the extended X-ray synchrotron flux 
   based on archival Chandra X-ray data, the analysis of which was presented by \citet{Nice2013}.
   The authors found no evidence of an X-ray PWN in the region around \mbox{PSR J1856+0245}
   and derived a $3\sigma$ upper limit on the unabsorbed $1-10$keV flux
   that translated into a limit on the luminosity of $\lessapprox 5 \times 10^{32}$ $\textrm{erg} \textrm{s}^{-1}$
   for an assumed distance of $9$ kpc.
   The extraction region used to derive this limit was an annulus centered on the pulsar
   with inner and outer radii of $2\arcsec$ and $15\arcsec$, respectively.
   This extended X-ray flux upper limit implies a weak magnetic field in the case of leptonic scenarios,
   with best-fit values in the models of \citet{Rousseau2012} being in the order of a few $\mu$G.

   We present our follow-up study of
   the complex morphology of \mbox{HESS J1857+026}.
   Section 2 describes the observations and data analysis performed using MAGIC.
   We summarize the MAGIC results in Section 3. In Section 4 we describe
   our multi-wavelength study of this region,
   which makes use of archival data in the radio and infrared bands.
   In Section 5 we present our physical interpretation of the
   VHE gamma-ray and multi-wavelength results.
   We summarize our conclusions in Section 6.
   

\section{Analysis with MAGIC}

   MAGIC is a system of two Imaging Atmospheric Cherenkov 
   Telescopes (IACTs) located at the Observatorio Roque de los Muchachos on the Canary Island of La Palma, Spain, 
   which has been operating in stereoscopic mode since 2009.
   Prior to this, MAGIC comprised a single IACT that began operating in 2004.
   MAGIC achieves a low trigger threshold of $50-60$ GeV, 
   while its sensitivity to point-like sources
   is 0.8 Crab units (C.U.) above 290 GeV in 50 hours \citep{Aleksic2012}. 
   
   We observed \mbox{HESS J1857+026} during July-October 2010, 
   taking 29 hours of data, for zenith angles between $25^{\circ}-36^{\circ}$.
   The data were taken in wobble mode by pointing the telescopes
   at four different pairs of pointing directions symmetric with respect to the source position,
   in order to obtain a more uniform and flat exposure. Two of these pairs
   were chosen to be at $0.4^{\circ}$ distance from the direction
   RA: $18^{h}57^{m}27^{s}$ and DEC: $02^{\circ}42'60''$,
   while the other two were $0.5^{\circ}$ away.

   The results presented here were obtained using the \emph{MARS} analysis framework \citep{Moralejo2009} 
   and using the image `sum-cleaning' algorithm \citep{Lombardi2011},
   which was found to improve the performance at lower energies, close to 100 GeV.
   The calibrated photomultiplier signals were cleaned
   to reduce the effect of the night sky background light,
   and the resulting image shape and timing information
   were combined between the two telescopes for each stereoscopic event.
   The gamma/hadron separation and the event direction reconstruction
   both made use of the random forest method \citep{Albert2008}.
   The estimated energy of a given event
   was determined from the brightness of the
   shower images, its reconstructed impact parameter and its incidence direction,
   using Monte Carlo (MC) filled look-up tables.
   Given that \mbox{HESS J1857+026} is an extended source
   our analysis used random forests and look-up tables that
   were trained using diffuse MC gamma-ray events.

   The determination of excess events in the signal region
   accounts for geometric exposure inhomogeneities
   by computing the background from the corresponding
   wobble partner data set and extracting it at the
   same relative camera plane coordinates as the signal region (further details in \citet{krause2013}).
   Our skymapping procedure models the background directly from the data
   and in bins of azimuth, thus taking azimuthal
   dependencies of the off-axis exposure into account \citep{Lombardi2011}.
   To compute the flux, we estimated the average effective area
   using a diffuse MC gamma-ray data sample
   selected as an annulus with radii $0.25^{\circ}$ and $0.65^{\circ}$,
   in order to account for variations in acceptance across the extent of the source.
   The resulting spectra were cross-checked in terms of the normalization
   of the estimated background inside the signal region by event numbers
   or by effective on-time with respect to the off-source regions.
   Furthermore, the unfolded spectrum was checked against
   four different spectral unfolding algorithms \citep{Albert2007},
   including the so-called \emph{forward folding} method,
   which fits the assumed spectrum by folding it with the energy response matrix
   and comparing the resulting distribution with the measured distribution of excess events.

   \section{Spectral and morphology results}
   
   \begin{figure*}
     \centering
     \includegraphics[height=.35\textheight]{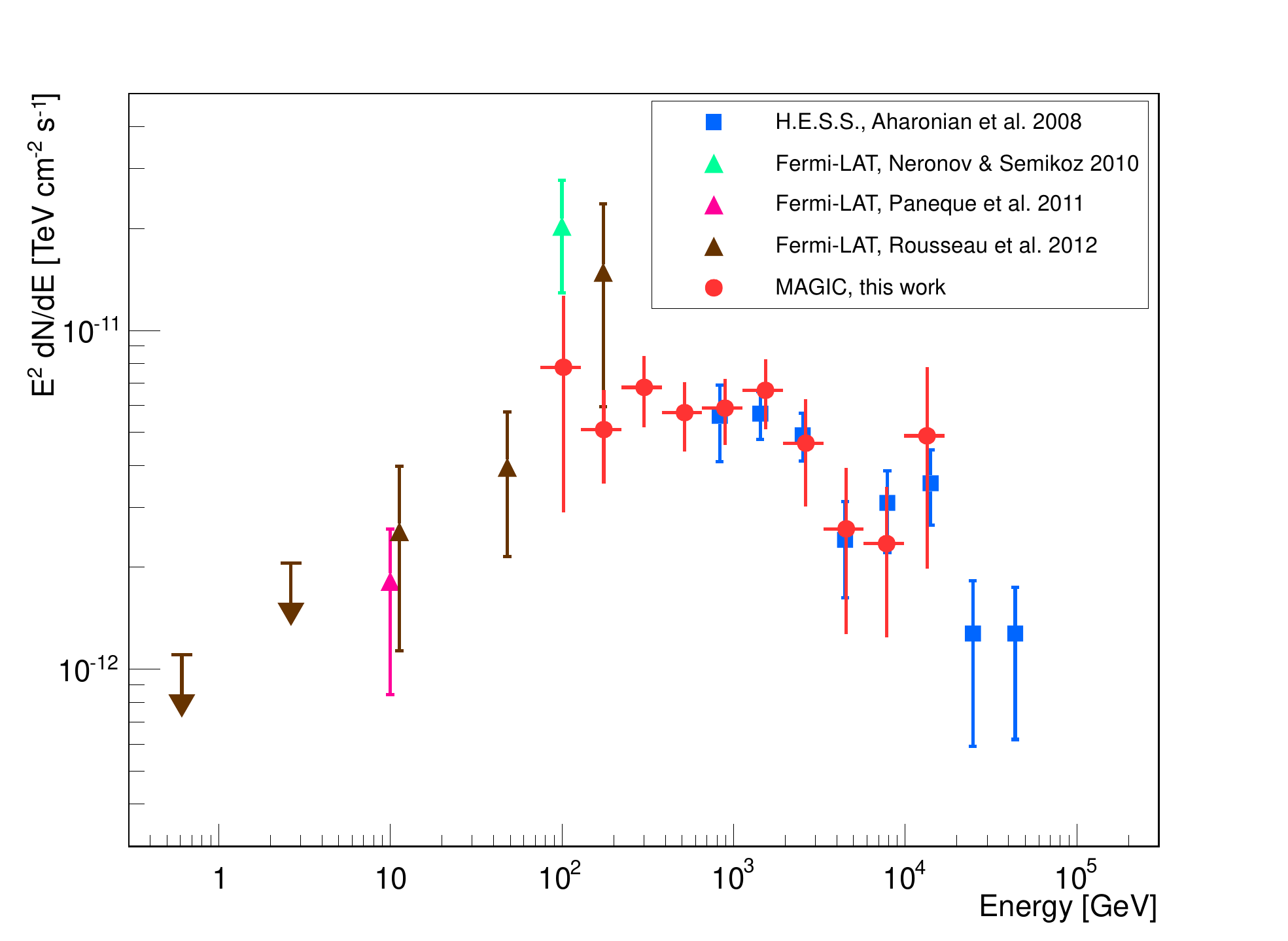}
     \caption{
       Spectral energy distributions of \mbox{HESS J1857+026} measured 
       by MAGIC (this work), H.E.S.S \citep{Aharonian2008}
       and \textit{Fermi}-LAT \citep{NeronovSemikoz2010,Paneque2011,Rousseau2012}.
       The MAGIC data are unfolded to correct for migration and energy biasing 
       effects \citep{Albert2007}, therefore the errors are not independent.
     }
     \label{fig1}
   \end{figure*}
   
   \begin{figure*}
     \centering
     \includegraphics[height=.33\textheight]{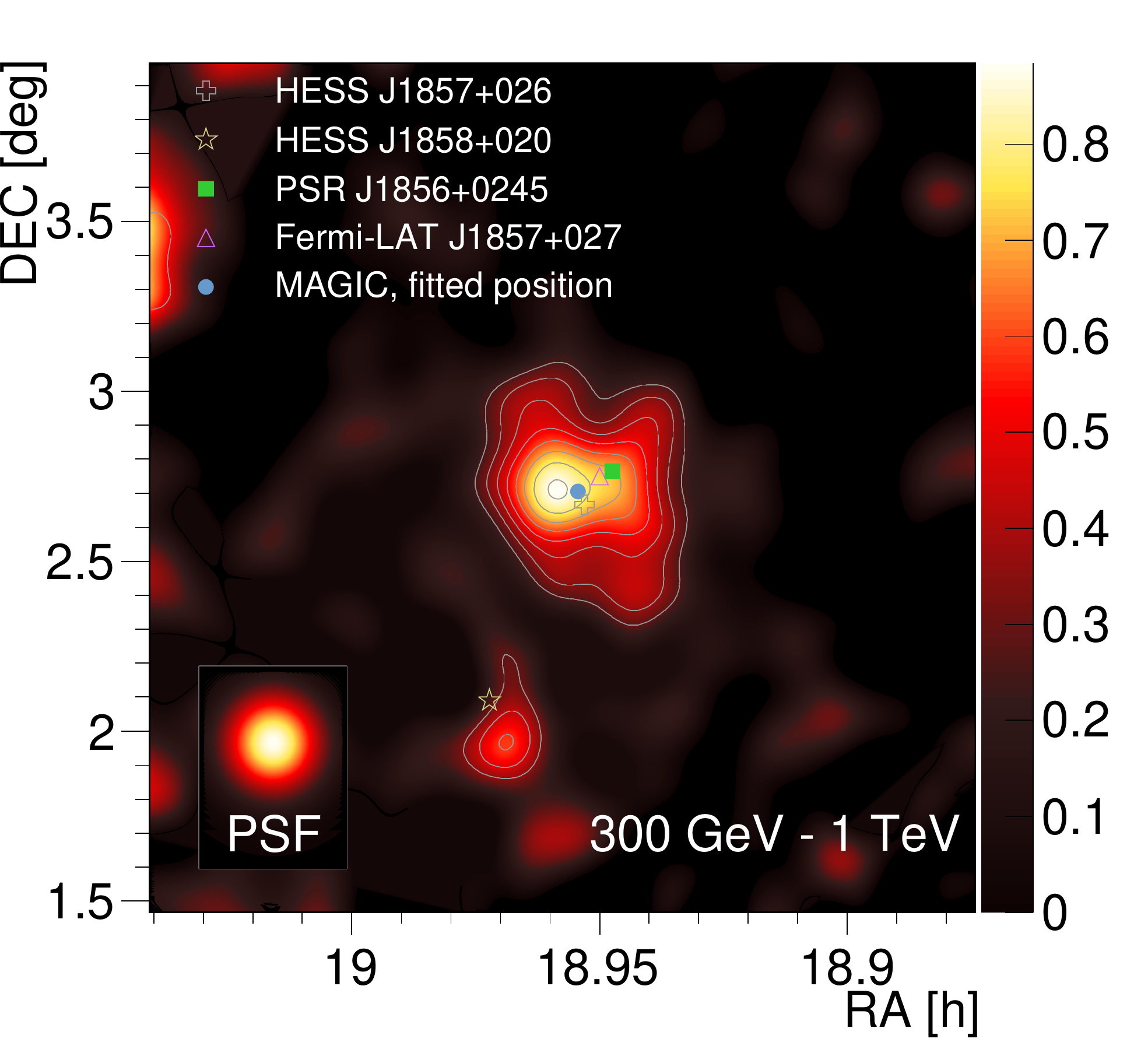}
     \includegraphics[height=.33\textheight]{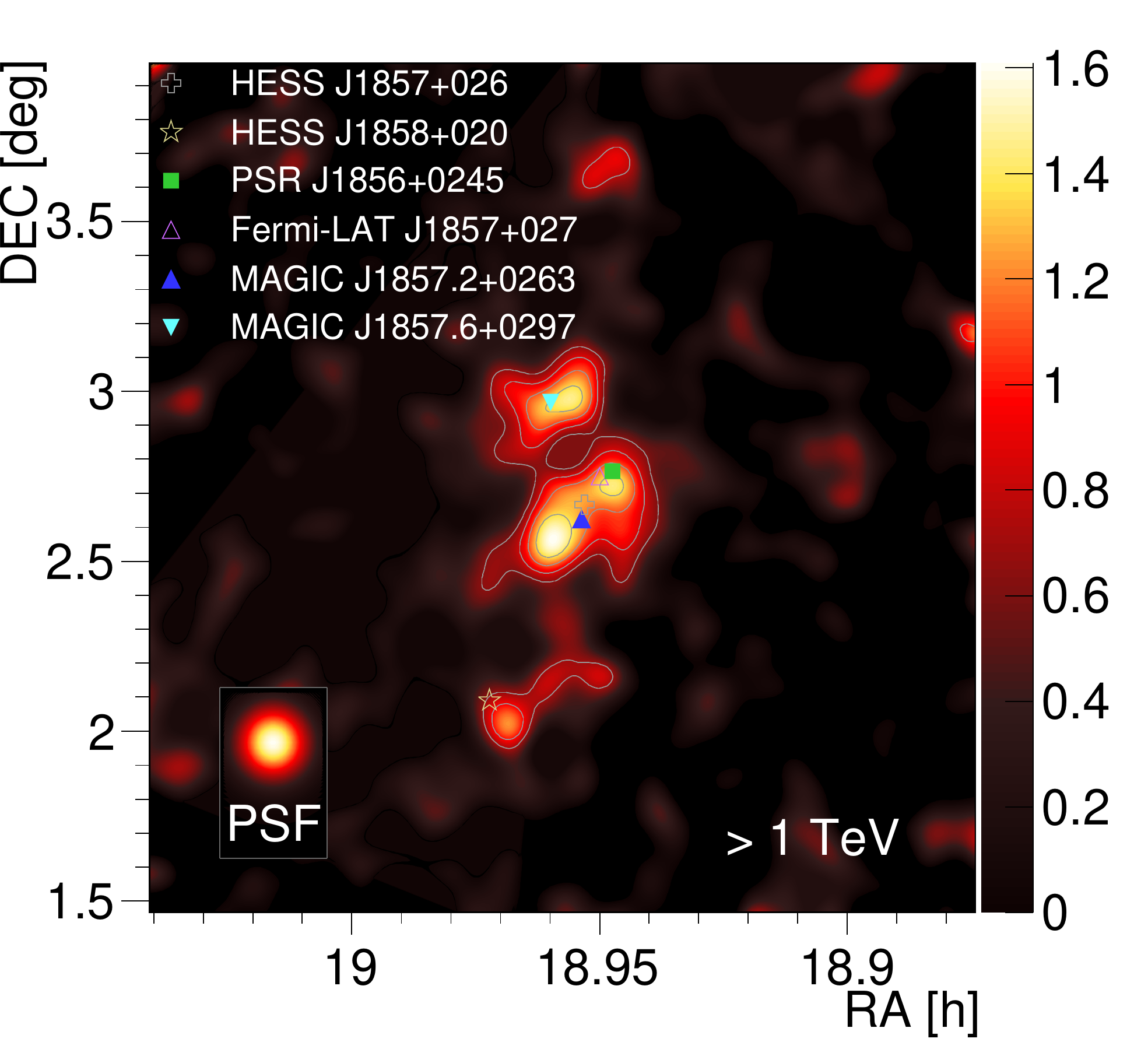}
     \caption{
       MAGIC gamma-ray flux map in arbitrary units (a.u.) for events with estimated energy
       300 GeV < E$_{\textrm{est}}$ < 1 TeV (left) and E$_{\textrm{est}}$ > 1 TeV (right).
       The median energies of these maps are estimated to be around 480 GeV and 1.7 TeV, respectively.
       The gamma-ray flux in arbitrary units
       is calculated as the number of smeared excess events in units of the background flux
       within $0.1^{\circ}$. Overlaid are test statistic (TS) value contours in steps of 1, starting at 3.
       They roughly correspond to Gaussian significances. Also shown in the bottom-left corner
       is the instrumental point spread function (PSF) after the applied smearing.
       The \textit{Fermi}-LAT source position is that determined by \citet{Rousseau2012}.
       The markers labeled \mbox{HESS J1857+026} and \mbox{HESS J1858+020}
       are centroid positions determined by \citet{Aharonian2008}.
     }
     \label{fig2}
   \end{figure*} 
   
   The MAGIC detection and spectrum of \mbox{HESS J1857+026}
   were obtained using a large circular signal extraction region with a radius of $0.4^{\circ}$,
   chosen to cover the entire extent of the source.
   Coupling this with a set of loose background rejection cuts with an MC gamma-ray efficiency of $95\%$,
   we detected the source above $\thicksim 150$ GeV
   with a significance of $12 \sigma$, using Eq. 17 of \citet{LiMa1983}.
   The spectrum measured by MAGIC was fitted through forward folding
   and is well described ($\chi^{2}/\textrm{d.o.f.} = 9.0/11$) by a single power law
   of the form $dN/dE = N_0(E/1\textrm{TeV})^{-\Gamma}$ with index 
   $\Gamma = 2.16\pm0.07_{stat} \pm0.15_{sys}$ and differential flux at 1 TeV
   of $N_0 = (5.37 \pm0.44_{stat} \pm1.5_{sys}) \times 10^{-12} (\textrm{TeV}^{-1} \textrm{cm}^{-2}$ $\textrm{ s}^{-1})$.
   The MAGIC spectrum is shown in \mbox{Figure \ref{fig1}}
   together with \textit{Fermi}-LAT \citep{NeronovSemikoz2010,Paneque2011,Rousseau2012}
   and H.E.S.S. \citep{Aharonian2008} results.
   Our spectral points have been unfolded to correct for migration and energy biasing 
   effects \citep{Albert2007}, therefore the errors are not independent.
   The measurement made by MAGIC connects those by H.E.S.S. and \textit{Fermi}-LAT,
   providing a continuous coverage of the spectral turnover close to 100 GeV.
   
   The gamma-ray flux sky maps in \mbox{Figure \ref{fig2}} show the energy-dependent morphology of \mbox{HESS J1857+026}.
   Based on MC gamma rays weighted using a spectral index of 2.3,
   the median energies of the low and high energy maps are around 480 GeV and 1.7 TeV, respectively.
   The source \mbox{HESS J1858+020} is not investigated here,
   given our relatively low exposure at its angular distance from the MAGIC pointing positions.
   In the estimated energy range of $0.3-1$ TeV,
   we have an instrument point spread function (PSF) of $0.079^{o}$
   and applied a smearing kernel of $0.077^{o}$
   such that the total PSF is $0.11^{o}$,
   while above $1$ TeV, our instrument PSF of $0.062^{o}$ and smearing kernel of $0.05^{o}$
   resulted in a total PSF of $0.08^{o}$.
   Each of the aforementioned numbers corresponds to the $1\sigma$ value of a symmetric 2D Gaussian function.
   We fitted a symmetric 2D Gaussian function to the $0.3-1$ TeV gamma-ray flux map
   and obtained a centroid position of
   RA: $18^{h}57^{m}15.7^{s} \pm5.8^{s}_{stat} \pm7.2^{s}_{sys}$ and DEC: $02^{\circ}42'17'' \pm1'26''_{stat} \pm1'48''_{sys}$
   which is compatible with the H.E.S.S. centroid position.
   We also measured an intrinsic source extension of $(0.20 \pm0.03_{stat} \pm0.02_{sys})^{\circ}$ from the fit,
   after removing the combined effect of the instrument PSF and the applied smearing.
   
   Above $1$ TeV we found that the VHE emission is due to two spatially distinct
   statistically significant components, which we denote \mbox{MAGIC J1857.2+0263}
   and \mbox{MAGIC J1857.6+0297}, as indicated in \mbox{Figure \ref{fig2}}.
   Using integration radii of $0.21^{\circ}$ for \mbox{MAGIC J1857.2+0263} and $0.14^{\circ}$ for \mbox{MAGIC J1857.6+0297},
   we computed significances of $6.7 \sigma$ and $6.0 \sigma$ respectively, by applying Eq. 17 of \citet{LiMa1983}.
   These radii were chosen to be as large as possible without having any overlap.
   In order to measure the positions and intrinsic extensions of the two components,
   we applied an iterative procedure in which we fitted one peak at a time
   while including the adjacent previously fitted 2D Gaussian function into the background.
   The elongated source \mbox{MAGIC J1857.2+0263} was fitted
   by a non-circular 2D Gaussian which yielded a centroid position
   RA: $18^{h}57^{m}13.0^{s} \pm4.0^{s}_{stat} \pm10.8^{s}_{sys}$ and DEC: $02^{\circ}37'31'' \pm50''_{stat} \pm3'_{sys}$
   and intrinsic extensions of $(0.17 \pm0.03_{stat} \pm0.02_{sys})^{\circ}$
   and $({0.06 \pm0.03_{stat} \pm0.02_{sys}})^{\circ}$ along its major and minor axes, respectively,
   with the major axis having an inclination of $(37\pm6_{stat})^{\circ}$ counter-clockwise with respect to the RA-axis.
   \mbox{MAGIC J1857.6+0297} was fitted by a circular 2D Gaussian which gave a mean position:
   RA: $18^{h}57^{m}35.6^{s} \pm5.3^{s}_{stat} \pm10.8^{s}_{sys}$ and DEC: $02^{\circ}58'02'' \pm56''_{stat} \pm3'_{sys}$
   and was compatible with a point source.
   
   We also tried to extract spectra around \mbox{MAGIC J1857.2+0263} and \mbox{MAGIC J1857.6+0297}
   using integration radii of $0.21^{\circ}$ and $0.14^{\circ}$, respectively.
   The limited event statistics above $1$ TeV and the inability to resolve the two peaks at lower energies
   made it difficult to extract spatially-resolved spectra.
   We estimate that the respective differential fluxes of \mbox{MAGIC J1857.2+0263} and \mbox{MAGIC J1857.6+0297}
   are $\thicksim 45\%$ and $\thicksim20\%$ of the total differential flux at $1$ TeV,
   which was measured from a circular region with of radius of $0.4^{\circ}$ and centered on RA: $18^{h}57^{m}27^{s}$ and DEC: $02^{\circ}42'60''$.
   We also determined a spectral index of $\Gamma = 2.2\pm0.1_{stat}$ for \mbox{MAGIC J1857.2+0263},
   which is compatible with that obtained for the entire region.
   
\section{Multi-wavelength view}
\label{sec:mwl}

  While we expect \mbox{MAGIC J1857.2+0263} to be associated with \mbox{PSR J1856+0245},
  the origin of the VHE emission from \mbox{MAGIC J1857.6+0297} is unidentified and
  we analyzed archival multi-wavelength data to look for possible counterparts.
  At radio wavelengths we used the 21 cm continuum and atomic hydrogen (H{\sc i}) line emission data from
  the Very Large Array (VLA) Galactic Plane Survey (VGPS) \citep{Stil2006}.
  At mm wavelengths we used the \mbox{$^{13}\textrm{CO}(\textrm{J}=1\rightarrow0)$} molecular line emission data from the Galactic ring survey (GRS) \citep{Jackson2006}
  to trace the molecular hydrogen (H$_2$) in this region.
  Finally, we examined $8$ $\mu$m images
  from the \emph{Spitzer} Galactic Legacy Infrared Midplane Survey Extraordinaire (GLIMPSE) Legacy Project \citep{Benjamin2013}.
  Further details regarding these data sets and our analysis are given in Appendix A.

  \begin{figure*}
    \centering
    \includegraphics[height=.23\textheight]{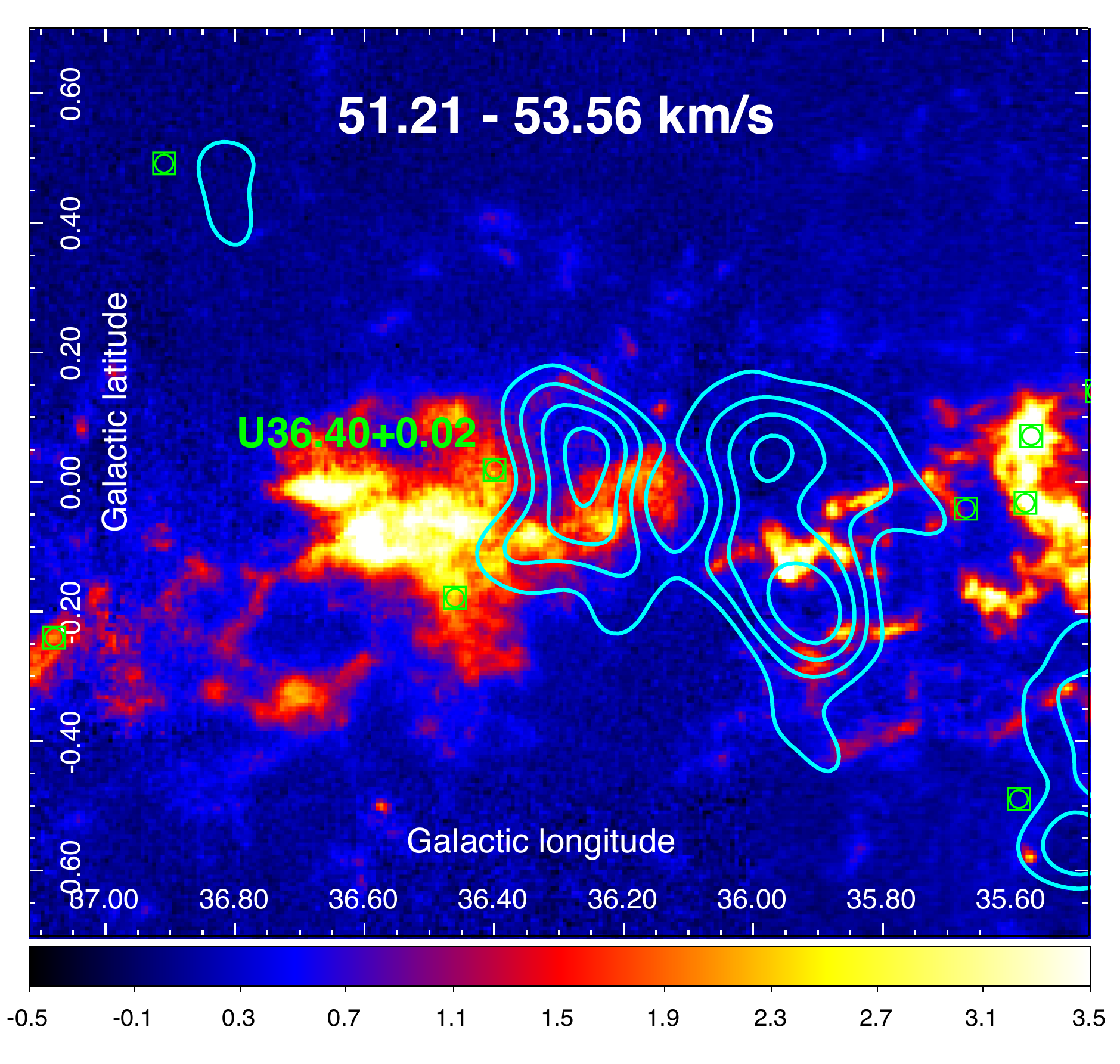}
    \includegraphics[height=.23\textheight]{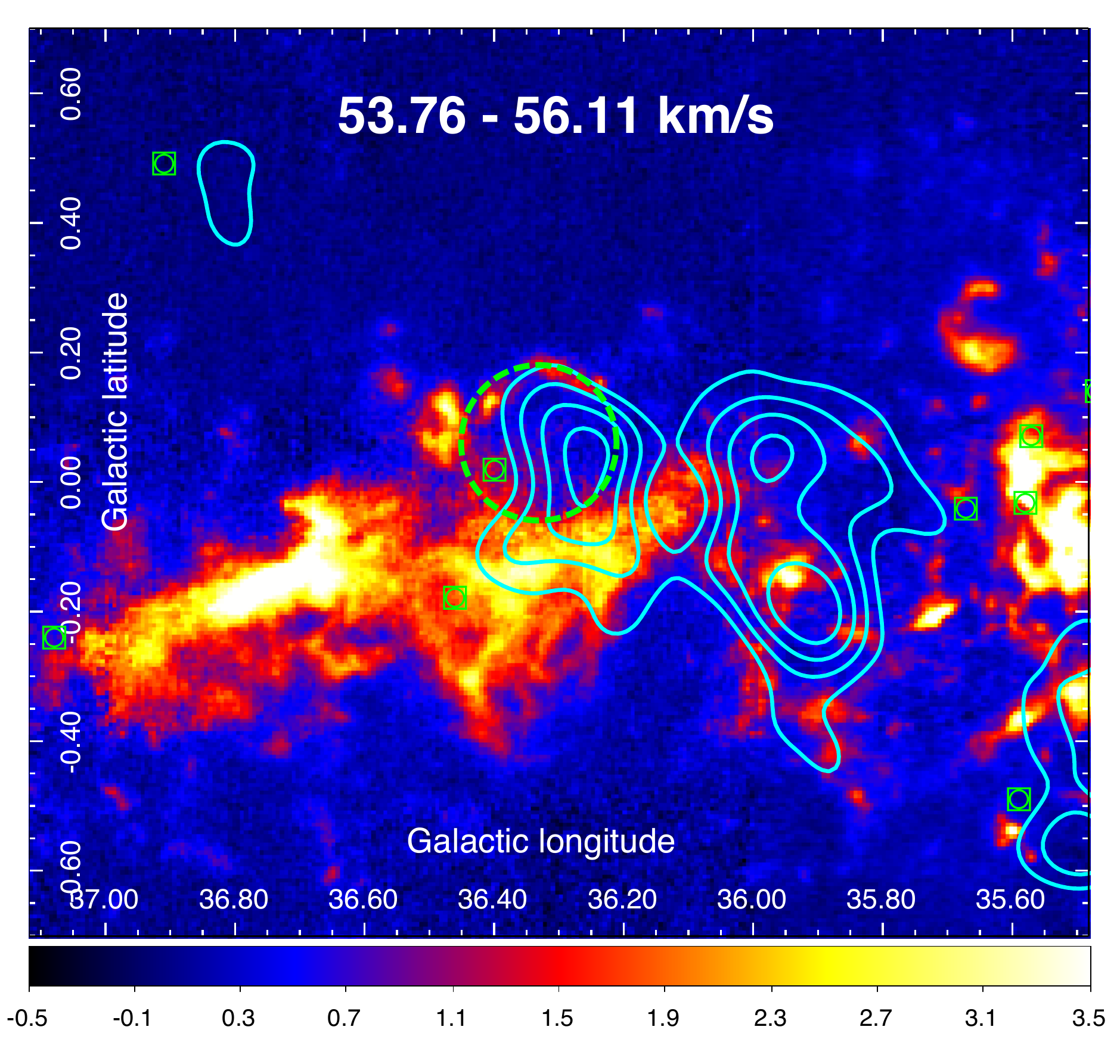}
    \includegraphics[height=.23\textheight]{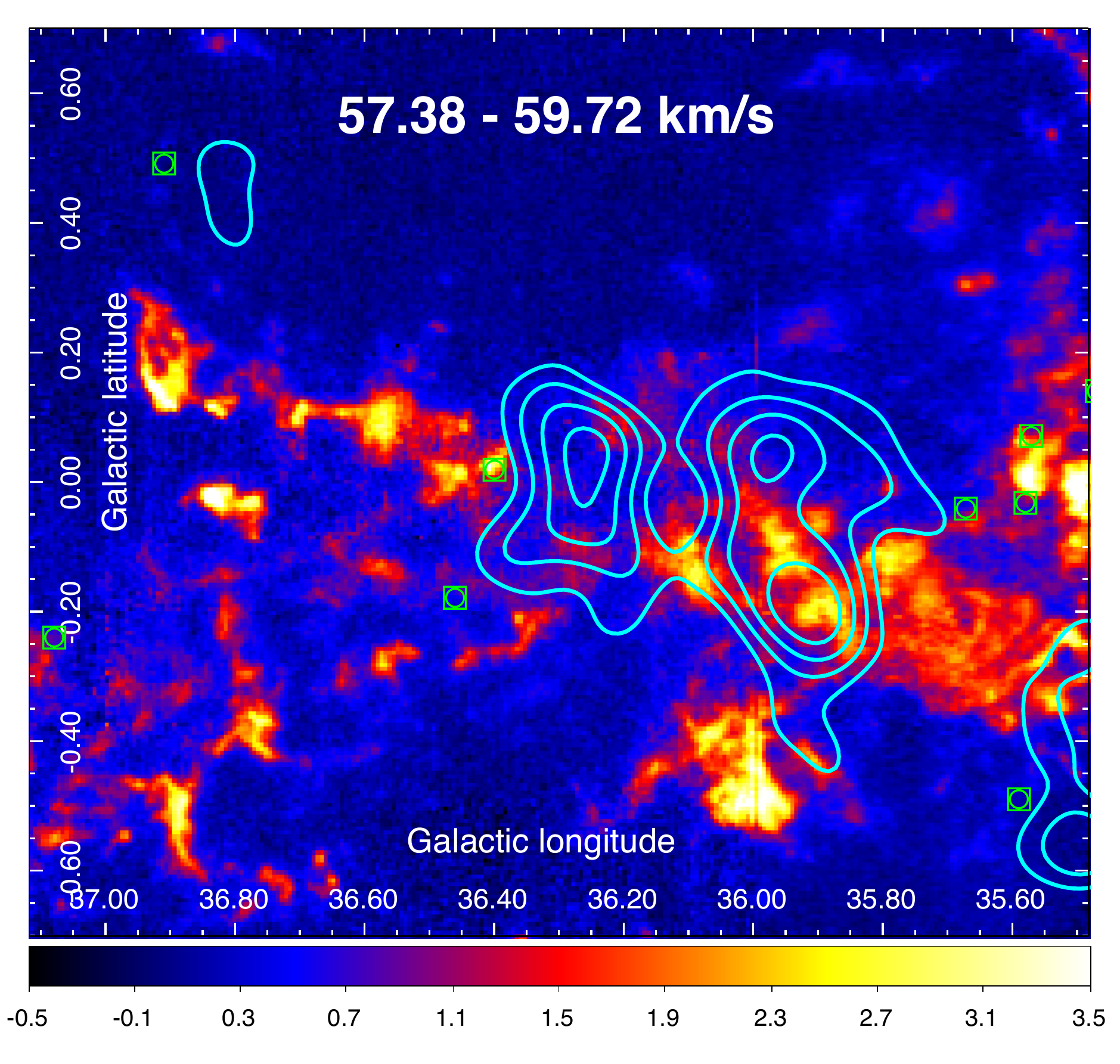}
    \caption{
      \mbox{$^{13}\textrm{CO}(\textrm{J}=1\rightarrow0)$} line emission integrated intensity (in units of K km s$^{-1}$) 
      in the vicinity of H{\sc ii} region \mbox{U36.40+0.02} ($V_{LSR} = 53.3 \textrm{km}\textrm{s}^{-1}$)
      and of the cataloged clouds G036.59-00.06 ($V_{LSR} = 53.49$ km s$^{-1}$)
      and G036.74-00.16 ($V_{LSR} = 55.19$ km s$^{-1}$),
      spanning three ranges of $V_{LSR}$:
      $51.21 - 53.56$ km s$^{-1}$ (left),
      $53.76 - 56.11$ km s$^{-1}$ (middle),
      $57.38 - 59.72$ km s$^{-1}$ (right).
      The cyan contours indicate the MAGIC $>1$ TeV TS levels in steps of 1, starting at 3.
      The circle-square green markers indicate H{\sc ii} regions in the catalog of \citet{Anderson2009}.
      The dashed green circle indicates the position of a possible cavity in the molecular gas.
    }
    \label{figco}
  \end{figure*}
  
  From the GRS data cube we computed the average
  \mbox{$^{13}\textrm{CO}(\textrm{J}=1\rightarrow0)$} emission spectrum
  over Galactic longitudes of $36.0^{\circ}$ to $37.0^{\circ}$
  and Galactic latitudes of $-0.5^{\circ}$ to $0.5^{\circ}$,
  finding two prominent peaks
  at kinematic local standard of rest (LSR) velocity ($V_{LSR}$) $\thicksim 55$ km s$^{-1}$ and $\thicksim 80$ km s$^{-1}$.
  We associate these peaks with three molecular gas clouds in the catalog of \citet{Roman-Duval2009}:
  G036.59-00.06, G036.74-00.16 and G036.49-00.16, with kinematic distances of $3.50$ kpc,  $3.62$ kpc and $8.38$ kpc, respectively.
  Most of the emission from G036.49-00.16 is outside the $\textrm{TS}=3$ contour of \mbox{MAGIC J1857.6+0297},
  which disfavors it as a counterpart.
  \mbox{Figure \ref{figco}} shows the molecular line emission toward \mbox{MAGIC J1857.6+0297}
  for a range of velocities that span those of G036.59-00.06 and G036.74-00.16.
  The range $51.21 - 53.56$ km s$^{-1}$ shows mainly the emission from G036.59-00.06,
  while at larger kinematic distances the range of $53.76 - 56.11$ km s$^{-1}$ shows emission from both G036.74-00.16 and G036.59-00.06.
  
  \begin{figure*}
    \centering
    \includegraphics[height=.2\textheight]{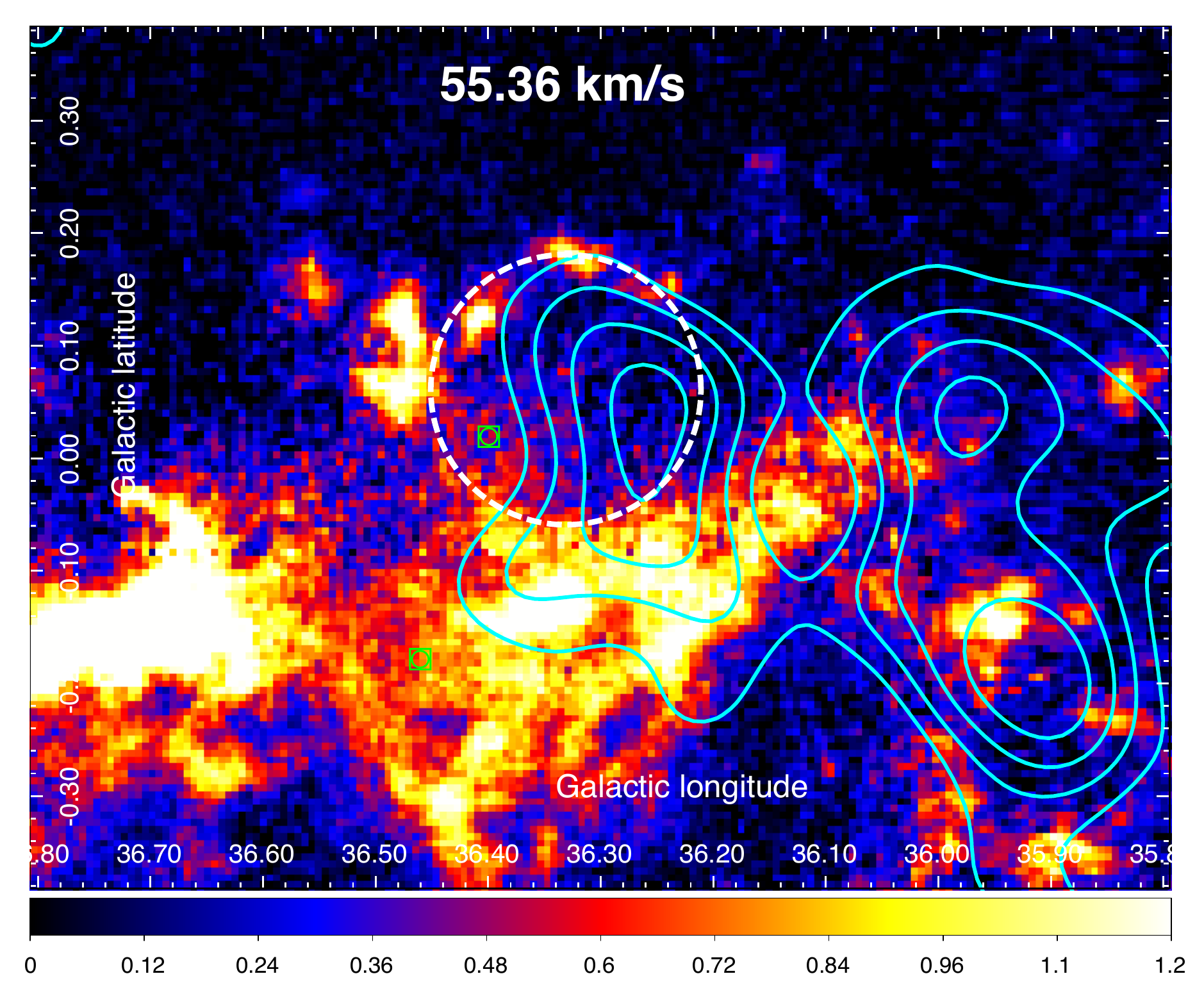}
    \includegraphics[height=.2\textheight]{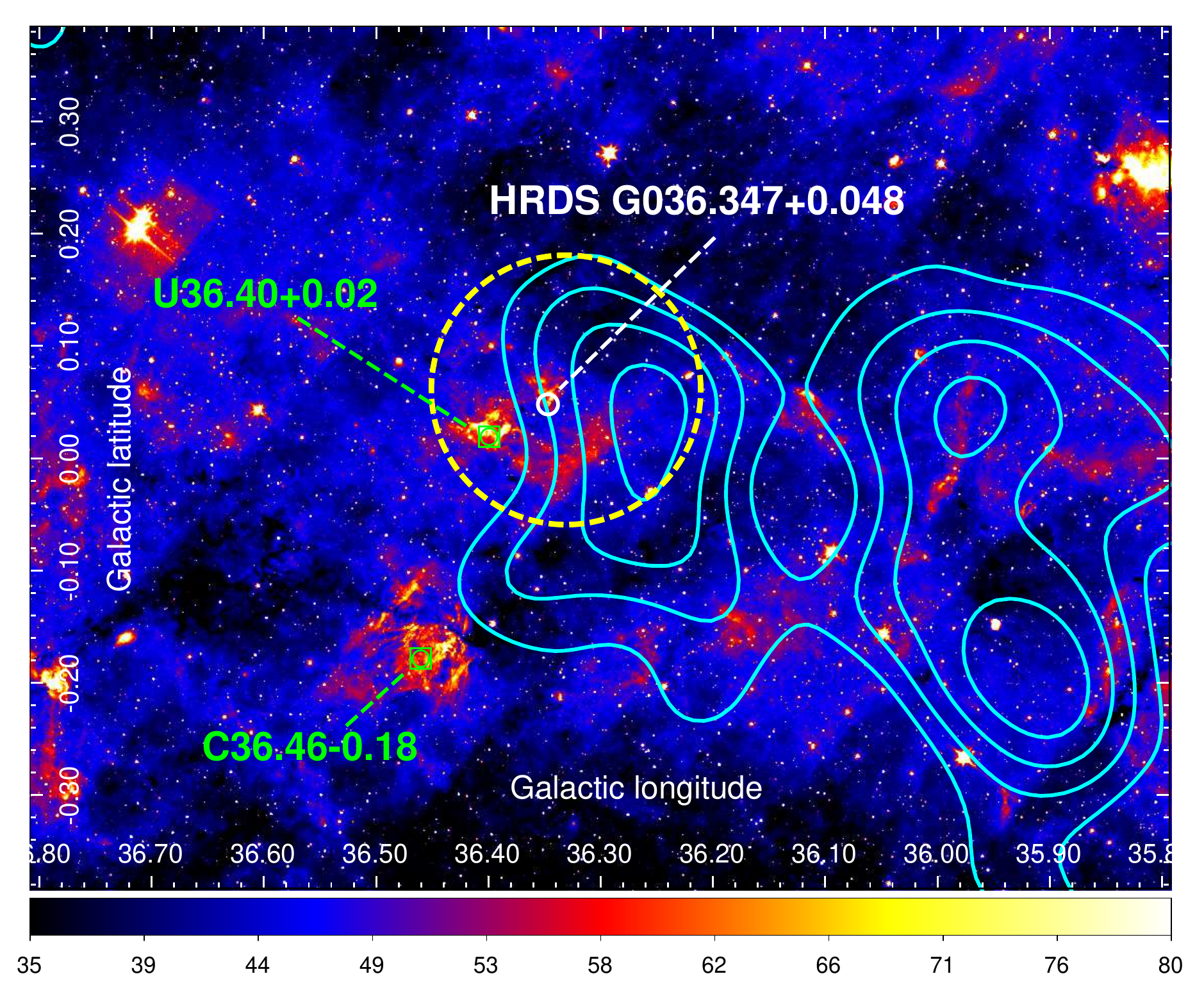}
    \includegraphics[height=.2\textheight]{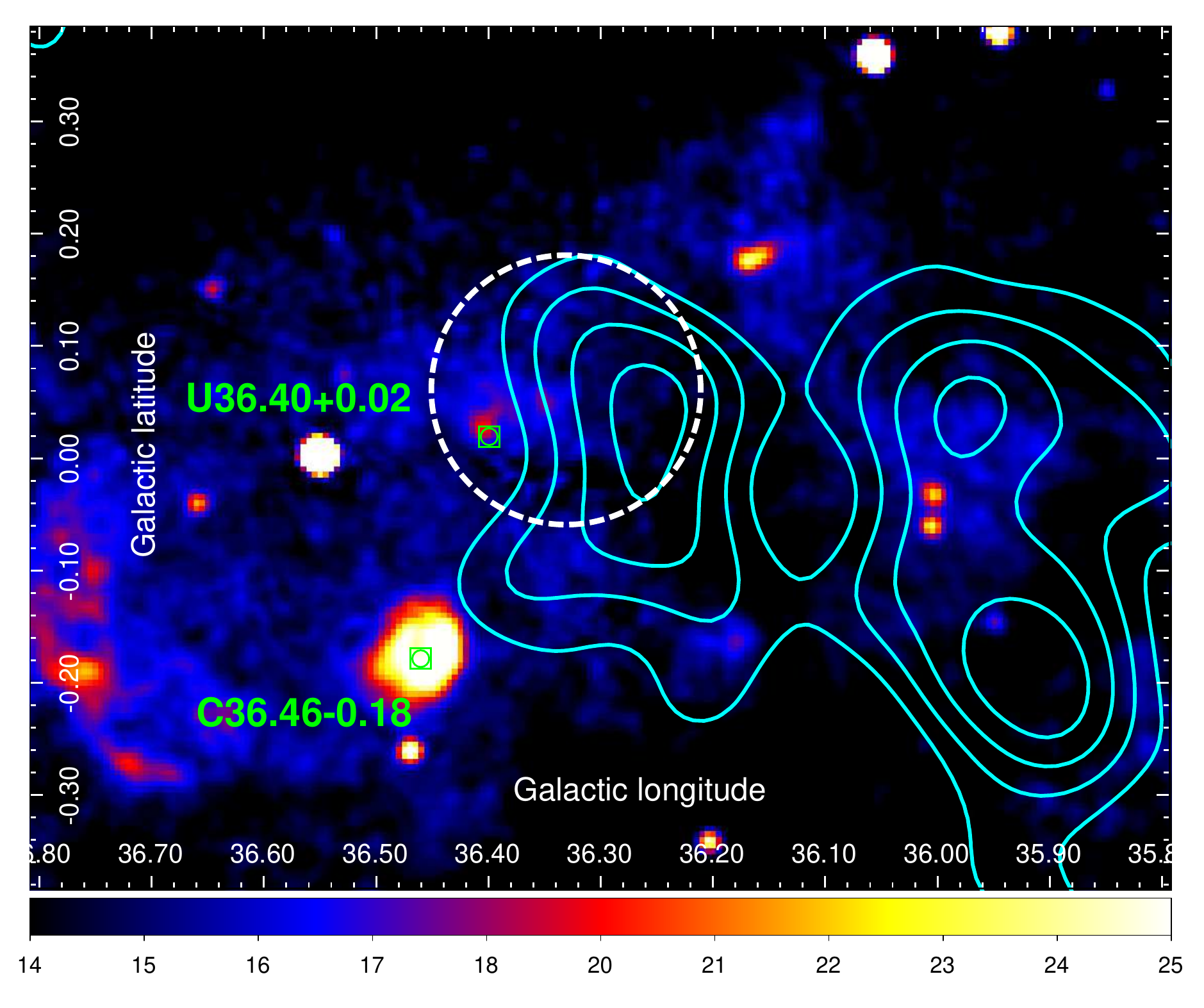}
    \caption{
      Left: \mbox{$^{13}\textrm{CO}(\textrm{J}=1\rightarrow0)$} line emission (in units of K)
      of a single velocity channel at $55.36$ km s$^{-1}$.
      Middle: GLIMPSE $8$ $\mu$m emission (in units of MJysr$^{-1}$).
      Right: VGPS $21$ cm continuum emission (in units of K).
      The cyan contours indicate the MAGIC $>1$ TeV TS levels in steps of 1, starting at 3.
      The green circle-square markers indicate H{\sc ii} regions in the catalog of \citet{Anderson2009},
      while the white circle marker indicates the H{\sc ii} region HRDS G036.347+0.048 \citep{Bania2012}.
      The larger dashed circle marks the position of a possible cavity in the molecular gas.
    }
    \label{figmwl}
  \end{figure*}
  
  Star formation takes place in the dense cores of massive molecular clouds,
  and this leads to a natural connection between molecular clouds and H{\sc ii} regions.
  The molecular gas properties of H{\sc ii} regions were explored by \citet{Anderson2009},
  and their kinematic distances were determined by \citet{AndresonBania2009} using the Galactic rotation curve of \citet{McClure-GriffithsDickey2007}.
  We searched these catalogs for H{\sc ii} regions close to the direction of \mbox{MAGIC J1857.6+0297}, finding two candidate counterparts.
  The first is the compact H{\sc ii} region \mbox{C36.46-0.18}, has a radio recombination line velocity of $V_{LSR} = 72.7 \pm 2.7$ km s$^{-1}$ \citep{Lokcman1989}
  and a fitted molecular gas velocity of $V_{LSR} = 75.99$ km s$^{-1}$, which corresponds to a distance of $9.0$ kpc \citep{AndresonBania2009}.
  Hence it is likely associated with the molecular cloud G036.49-00.16 ($V_{LSR} = 76.87$ km s$^{-1}$) and we consider it unrelated to the VHE emission.
  The second candidate is the ultra-compact H{\sc ii} region \mbox{U36.40+0.02}.
  Observations of H$110\alpha$ emission from this region
  determined a velocity of $V_{LSR} = 53.3 \pm 1.4$ km s$^{-1}$ and a corresponding near kinematic distance of $3.7^{+0.7}_{-0.6}$ kpc \citep{Watson2003},
  assuming the Galactic rotation curve of \citet{{BrandBlitz1993}}.
  H$_{2}$CO absorption line measurements by \citet{Watson2003} indicated the presence of molecular clouds
  at  $V_{LSR} = 52.5 \pm 0.2$ km s$^{-1}$ and at  $V_{LSR} = 57.9 \pm 0.2$ km s$^{-1}$.
  \mbox{U36.40+0.02} has a fitted molecular gas velocity of $V_{LSR} = 52.42$ km s$^{-1}$ with corresponding near kinematic distance of $3.3$ kpc \citep{AndresonBania2009},
  and hence is likely to be associated with the cloud G036.59-00.06 ($V_{LSR} = 53.49$ km s$^{-1}$).
  
  The velocity ranges of $51.21 - 53.56$ km s$^{-1}$ and $57.38 - 59.72$ km s$^{-1}$ in \mbox{Figure \ref{figco}}
  were selected to illustrate the two prominent \mbox{$^{13}\textrm{CO}$} emission peaks
  in the molecular gas complex surrounding \mbox{U36.40+0.02}.
  Between them, in the range of $53.76 - 56.11$ km s$^{-1}$, we discovered an incomplete shell-like structure
  coincident with the direction of \mbox{MAGIC J1857.6+0297}.
  Furthermore, the Galactic longitude-velocity and latitude-velocity plots in \mbox{Figure \ref{figlbv}} of Appendix B
  show a perturbed spatial and velocity structure in the side of the cloud G036.59-00.06,
  in a region that roughly corresponds to the directions and velocity channels over which this feature is visible.
  In particular, the latitude-velocity signature indicates that the partial shell spans $\thicksim 3$ km s$^{-1}$ in velocity space.
  \mbox{Figure \ref{figmwl}} shows a single GRS velocity channel at $55.36$ km s$^{-1}$ that provides a close-up view
  of what could be a gas cavity or wind-blown bubble.
  
  In \mbox{Figure \ref{figmwl}} we also show the 8 $\mu$m emission,
  which primarily traces emission from polycyclic aromatic hydrocarbons (PAHs) heated by UV starlight.
  These carbon molecules are found in photodissociation regions (PDRs),
  which interface the ionized matter in H{\sc ii} regions and the molecular gas of their parent clouds.
  The 8 $\mu$m emission from \mbox{U36.40+0.02}, which was originally called IRAS 18551+0302 \citep{Beichman1988},
  shows a possible diffuse component that extends toward the center of the proposed gas cavity.
  A second, more compact, component corresponds to the newly discovered H{\sc ii} region \mbox{HRDS G036.347+0.048} \citep{Bania2012}.
  While a kinematic distance has not been determined for this source, its hydrogen radio recombination
  line (Hn$\alpha$) spectrum yielded a $V_{LSR}$ of $76.9$ km s$^{-1}$, so it is not physically related to \mbox{U36.40+0.02}.
  A survey of IRAS sources using the high-density gas tracer \mbox{$\textrm{CS}(2\rightarrow1)$}
  provided a velocity of $V_{LSR} = 58.0$ km s$^{-1}$ for IRAS 18551+0302 \citep{Bronfman1996}.
  This measurement likely traces the gas clump seen in the $V_{LSR}$ range of $57.38 - 59.72$ km s$^{-1}$ of \mbox{Figure \ref{figco}}.
  Ammonia emission, which traces high density ($\thicksim 10^4$ cm$^{-3}$) gas, was also detected by the Red MSX Source (RMS) survey \citep{Urquhart2011},
  in which the source, \mbox{MSX6C G036.4057+00.0230}, had a NH$_3$ ($1$,$1$) velocity of $V_{LSR} = 57.77$ km s$^{-1}$.
  The authors also detected H$_2$O maser emission, which indicates on-going star formation.
  This was found over the velocity range $47.2 - 53.1$ km s$^{-1}$
  with the peak emission at $V_{LSR} = 48.4$ km s$^{-1}$,
  which is just below the $V_{LSR}$ range of $51.21 - 53.56$ km s$^{-1}$ shown in \mbox{Figure \ref{figco}}.
  
  \begin{figure*}
    \centering
    \includegraphics[height=.23\textheight]{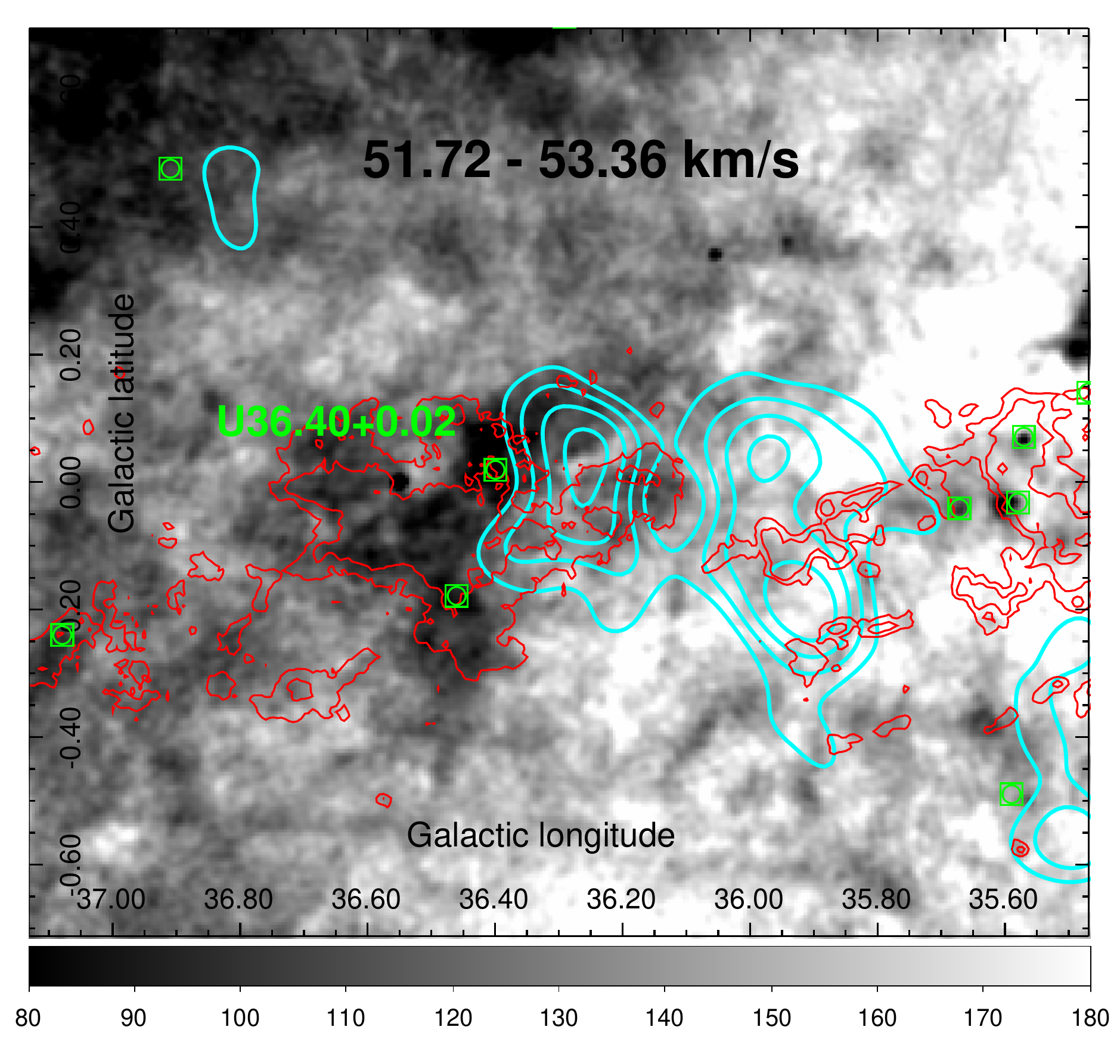}
    \includegraphics[height=.23\textheight]{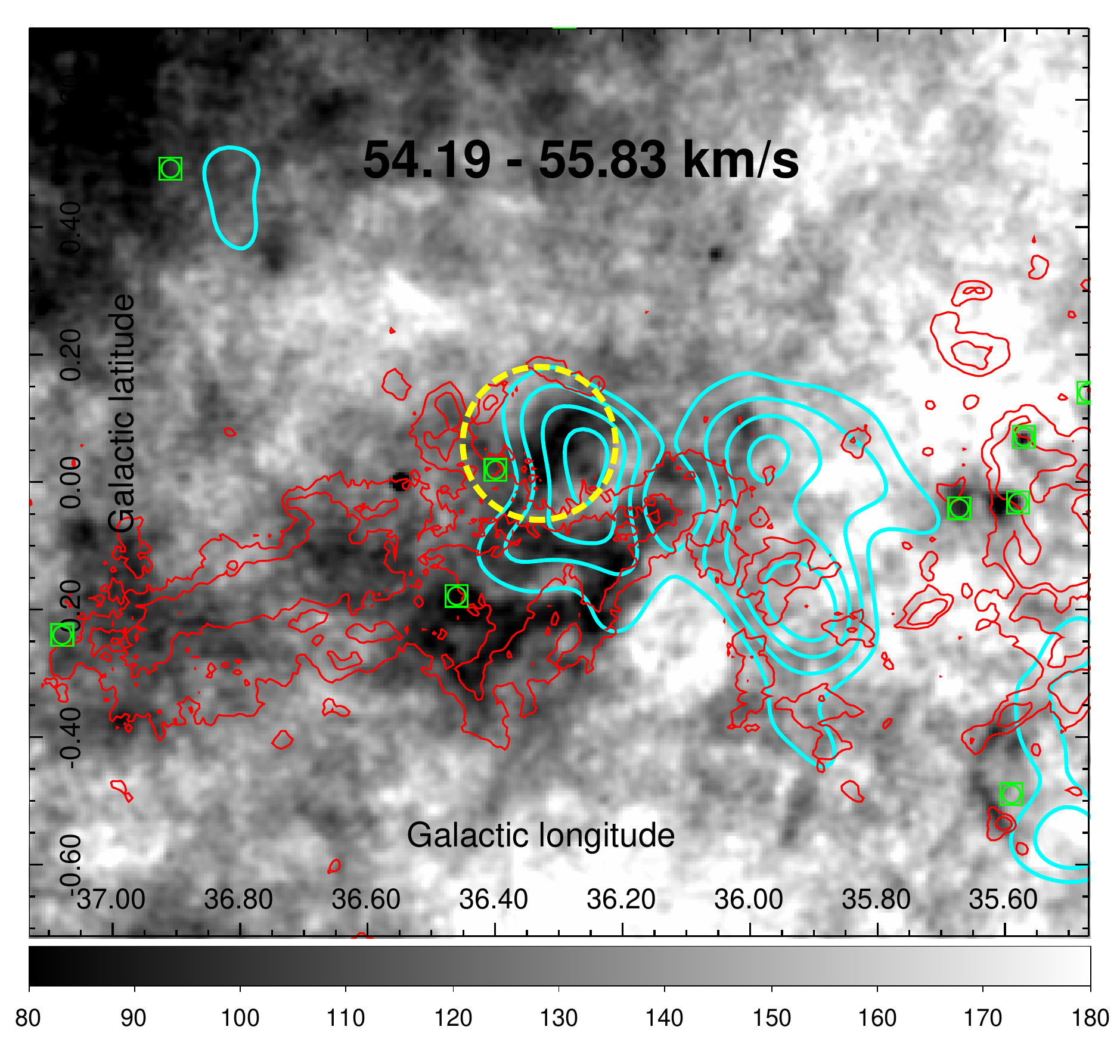}
    \includegraphics[height=.23\textheight]{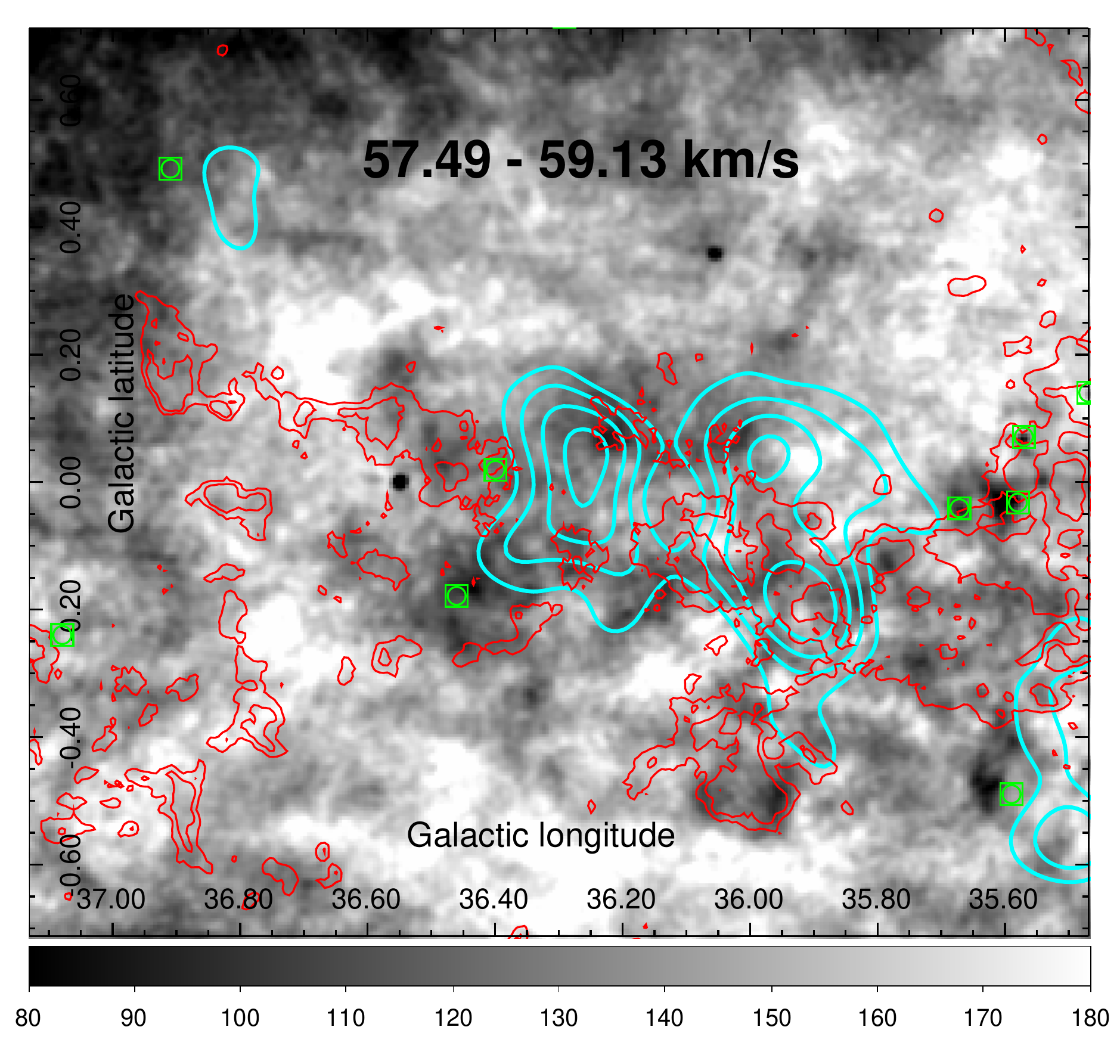}
    \caption{
      H{\sc i} line emission integrated intensity (in units of K km s$^{-1}$) 
      in the vicinity of H{\sc ii} region \mbox{U36.40+0.02} ($V_{LSR} = 53.3 \textrm{km}\textrm{s}^{-1}$)
      for three ranges of $V_{LSR}$:
      $51.72 - 53.36$ km s$^{-1}$ (left),
      $54.19 - 55.83$ km s$^{-1}$ (middle),
      $57.49 - 59.13$ km s$^{-1}$ (right).
      The red contours indicate \mbox{$^{13}\textrm{CO}(\textrm{J}=1\rightarrow0)$} integrated intensity levels of $1$ and $2$ K km s$^{-1}$
      from the corresponding velocity intervals in \mbox{Figure \ref{figco}}.
      The cyan contours indicate the MAGIC $>1$ TeV TS levels in steps of 1, starting at 3.
      The circle-square green markers indicate H{\sc ii} regions in the catalog of \citet{Anderson2009}.
      The dashed yellow circle indicates the position of a possible cavity in the molecular gas.
    }
    \label{fighi}
  \end{figure*}
  
  In \mbox{Figure \ref{fighi}} we display the H{\sc i} 21 cm line emission from the VGPS
  over three velocity ranges that approximately correspond to those of \mbox{Figure \ref{figco}}.
  The most compact  H{\sc i} absorption features,
  produced against strong background continuum emission \citep{KucharBania1990},
  can be either galactic (e.g. H{\sc ii} regions) or extragalactic in origin.
  The velocity ranges $51.72 - 53.36$ km s$^{-1}$ and $54.19 - 55.83$ km s$^{-1}$
  exhibit absorption from cold neutral atomic hydrogen that is associated with a molecular cloud
  absorbing against more distant warm 21 cm emitting gas (e.g. see \citet{Roman-Duval2009}).
  Cold, optically thick H{\sc i} can form in the dense parts of molecular clouds through the dissociation of $H_{2}$ by penetrating cosmic rays.
  \citet{GoldsmithLi2005} found that its resulting self-absorption features
  tend to correlate with peaks in the \mbox{$^{13}\textrm{CO}(\textrm{J}=1\rightarrow0)$} emission.
  We note, however, that the dark region located within the yellow dashed circle shown in \mbox{Figure \ref{fighi}}
  does not spatially coincide with any bright \mbox{$^{13}\textrm{CO}(\textrm{J}=1\rightarrow0)$} emission in the same velocity range,
  nor with any strong background continuum source (see \mbox{Figure \ref{figmwl}}).
  This feature may be explained as self-absorption by cold atomic gas components surrounding the molecular cloud.
  These are expected to form via the photodissociation of $H_{2}$ by interstellar UV photons \citep{Jackson2002}.
  Such optically thick H{\sc i} may also trace $H_{2}$ that is not detected in $^{13}\textrm{CO}$, 
  or alternatively, it may point to a strong shock induced depletion of the dust grains,
  which provide photoelectric heating and facilitate $H_{2}$ formation (see \cite{Gibson2010}, and references therein).
  Finally, the feature could also be a real void in the atomic gas.
  We note that the possible presence of OB stars would lead to a natural connection between such a cavity
  and the surrounding \mbox{$^{13}\textrm{CO}$} shell-like feature.

\section{Discussion}

   The broadband SED of HESS J1857+026 in \mbox{Figure \ref{fig1}} exhibits a strong turnover at energies 
   close to 100 GeV. This may be attributed to the IC peak 
   expected from a leptonic scenario, although \citet{Rousseau2012} showed that 
   the SED could also be modeled in terms of gamma-ray emission with a hadronic origin.
   Here we consider whether the proposed PWN scenario can adequately 
   explain the morphology of \mbox{HESS J1857+026} observed with HESS, MAGIC and \textit{Fermi}-LAT.

   While we find that the source is significantly extended
   with respect to our PSF in the energy range of $0.3-1$ TeV,
   no significant extension was found by \citet{Rousseau2012} using \textit{Fermi}-LAT data for energies above 10 GeV.
   The MAGIC $0.3-1$ TeV intrinsic extension of $0.20 \pm0.03_{stat}$ however,
   is comparable to the size of the instrument PSF  ($68$\% containment radius) of \textit{Fermi}-LAT above 10 GeV,
   and this may explain the non-detection of an intrinsic extension for this source using \textit{Fermi}-LAT.
   Indeed \citet{Acero2013} recently specified a 99\% confidence level upper limit of $0.28^{\circ}$ on the extension of this source.
   Comparing with H.E.S.S., the intrinsic extension measure of $(0.11 \pm 0.08_{stat})^{\circ} \times (0.08 \pm 0.03_{stat})^{\circ}$
   in their energy range is consistent with our $0.3-1$ TeV extension measurement.
   Furthermore, if we only attribute the PWN emission above 1 TeV to \mbox{MAGIC J1857.2+0263},
   its size of $(0.17 \pm 0.03_{stat})^{\circ} \times (0.06 \pm 0.03_{stat})^{\circ}$
   is in good agreement with the H.E.S.S. result,
   although it is not clear to what extent the north tail-like emission,
   which is seen in the H.E.S.S. skymap and may correspond to \mbox{MAGIC J1857.6+0297}, influenced their 2D Gaussian fit.

   By itself, the MAGIC energy-dependent morphology would favor a leptonic PWN scenario
   in which lower energy electrons can diffuse out to larger distances from the pulsar,
   but only if we consider \mbox{MAGIC J1857.2+0263} in the $>1$ TeV map of \mbox{Figure \ref{fig2}}
   as an independent source that becomes less compact at $0.3-1$ TeV.
   This implies that the fainter \mbox{MAGIC J1857.6+0297} emission,
   which we only resolve above $1$ TeV, is unrelated to \mbox{PSR J1856+0245}.
   This is because if \mbox{MAGIC J1857.6+0297} was due to the PWN electrons,
   its distance from the pulsar would be $\thicksim 43 \times (d/9\textrm{kpc})$ pc,
   where $d$ is the distance to \mbox{PSR J1856+0245},
   and this would make the entire emission region more extended above $1$ TeV than below.
   Moreover, if \mbox{MAGIC J1857.6+0297} did indeed trace the same PWN as \mbox{MAGIC J1857.2+0263},
   the apparent hole between these emission regions makes it less likely that both originate from
   the IC scattering of ambient photons on the same electron population.
   
   While we do not consider any sub-structure within \mbox{MAGIC J1857.2+0263} to be significant,
   we note that \mbox{PSR J1856+0245} is located close to one end of its major axis.
   Using the observed angular elongation of \mbox{MAGIC J1857.2+0263}
   we can estimate the physical length of the PWN emission region to be  $\thicksim 40 \times (d/9\textrm{kpc})$ pc.
   The VHE gamma-ray emission of a PWN traces a relic electron population accumulated over its lifetime.
   If we consider the characteristic spin-down age of \mbox{PSR J1856+0245} ($21$ kyr) and assume a typical velocity of $400-500$ km s$^{-1}$,
   the offset resulting from its proper motion would be at most $\thicksim10 \times (d/9\textrm{kpc})$ pc.
   This maximum value corresponds to the extreme case in which the velocity direction is perpendicular to the line of sight.
   However, as outlined by \citet{GaenslerSlane2006}, the morphology of middle-aged PWN
   can be strongly influenced by the complex interaction between the PWN and the reverse shock of its SNR.
   This can distort the PWN morphology and cause an offset in the position of the pulsar with respect to the center of VHE emission.
   Therefore the observed morphology  of \mbox{MAGIC J1857.2+0263} may indicate that such evolutionary effects
   have played some role in its elongated appearance.

   The nature of the VHE emission from \mbox{MAGIC J1857.6+0297} remains a mystery,
   although our targeted multi-wavelength study provides some clues to its origin.
   The shell-like morphology of the \mbox{$^{13}\textrm{CO}$} channel maps seen in Figures \ref{figco},
   coupled with the corresponding signatures of perturbed gas in Figure \ref{figlbv},
   points to the presence of a cavity or wind-blown bubble of radius $\thicksim 8 \times (d/3.7\textrm{kpc})$ pc
   that coincides with the direction of the VHE gamma-ray emission peak.
   Here we consider possible scenarios in which the proposed cavity and the nearby ultra-compact H{\sc ii} region \mbox{U36.40+0.02},
   which is embedded in this molecular gas complex, are linked to \mbox{MAGIC J1857.6+0297}.
   That the peak of the TeV emission does not spatially correspond to the $^{13}\textrm{CO}$ emission peaks
   in any of the velocity ranges shown in Figure \ref{figco} would argue against a hadronic origin,
   where molecular hydrogen acts as target matter for the accelerated hadrons.
   The emission could therefore have a leptonic origin, where accelerated electrons may
   IC scatter on the IR to UV photon fields of \mbox{U36.40+0.02}.
   We do not, however, exclude a hadronic origin due to the possible presence
   of cold, dense atomic gas within the proposed $^{13}\textrm{CO}$ cavity,
   which could also act as target material for accelerated hadrons (e.g. see \citet{Fukui2012}).
   
   A leptonic scenario that may explain the radio data
   has the VHE emission originating from a PWN
   whose progenitor star was in an OB association formed within \mbox{U36.40+0.02},
   the stellar winds of which created the cavity.
   As outlined in \citet{dejageratai2008}, this could then facilitate an initial unimpeded expansion of the PWN,
   allowing for weaker magnetic fields that would result in lower levels of synchrotron cooling
   for the VHE gamma-ray emitting electrons.
   A similar scenario was also proposed by \citet{bockgavaramadze2002} to explain
   the possible association of the pulsar PSR B1706-44 with SNR G343.1-2.3,
   which are candidate counterparts of the extended source HESS J1708-443 \citep{Abramowski2011}.
   \citet{bockgavaramadze2002} suggested that in this case,
   the massive progenitor star escaped its molecular cloud
   and produced a wind-blown bubble in the less dense ISM
   together with a wind-driven cavity in the parent cloud. They
   also speculated that the subsequent SN blast wave
   then interacted on one side with the shell of the bubble
   and with the parent cloud on the other side.
   In the case of \mbox{MAGIC J1857.6+0297},
   additional high-resolution multi-wavelength data at radio wavelengths
   and in X-rays will be necessary to fully test such a scenario.
   
   An alternative source of particle acceleration to explain the observed VHE emission
   could be the combined effect of the strong stellar winds themselves.
   For instance, VHE gamma rays may be produced by particles
   accelerated in the colliding winds of early-type
   (O, early B, Wolf-Rayet) star binary systems,
   although \citet{Reimer2006} showed that the spectra of such systems
   are expected to suffer from strong absorption above $\thicksim50$ GeV
   due to pair production in the surrounding photon fields.
   Given that \mbox{U36.40+0.02} and the surrounding dense molecular gas
   host regions of on-going star formation, as evidenced by the H$_2$O maser emission,
   the VHE emission could instead stem from particles accelerated to multi-TeV energies
   at the shocks produced by outflows from massive protostars \citep{araudo2008,bosch-ramon2010}.
   Since the kinetic luminosity of one such jet is expected to be around $10^{36} \textrm{erg} \textrm{s}^{-1}$,
   a collection of these objects would also be required to explain the VHE emission.

\section{Conclusions}
   
   MAGIC has performed observations of the TeV PWN candidate source \mbox{HESS J1857+026}
   in order to better understand its complex morphology.
   We have also extended the VHE spectrum of the source down to 100 GeV,
   thus bridging the gap between the earlier spectral measurements of H.E.S.S.
   and those obtained more recently with \textit{Fermi}-LAT.
   Assuming that the observed VHE gamma-ray emission is produced by energetic electrons via the IC channel,
   their energy-dependent diffusion would dictate a more extended PWN about \mbox{PSR J1856+0245}
   when going down in energy from the multi-TeV regime to the range covered by \textit{Fermi}-LAT.
   Our study of the energy-dependent morphology
   has revealed for the first time that this is is not the
   case, with the source becoming more extended above 1 TeV.
   Based on this finding we have proposed a two-source scenario for
   the separated and statistically significant peaks in emission that we detected above 1 TeV,
   which we have called \mbox{MAGIC J1857.2+0263} and \mbox{MAGIC J1857.6+0297}.
   We then interpret \mbox{MAGIC J1857.2+0263} as the emission from the relic PWN of \mbox{PSR J1856+0245}.
   We expect that the elongated shape of the emission above 1 TeV 
   as well as its offset position from the pulsar
   are signatures of complex PWN evolutionary processes,
   which involve an interaction with the SNR reverse shock,
   while the initial pulsar velocity could also play a role.
   
   We have used archival multi-wavelength data
   at radio and infra-red wavelengths to search for possible
   counterparts of \mbox{MAGIC J1857.6+0297}. We found a possible
   association with an ultra-compact H{\sc ii} region located at $3.7$ kpc
   and its neighboring molecular clouds.
   The morphology of molecular gas in this region suggests that a cavity
   or wind-blown bubble may be present just behind the H{\sc ii} region 
   and in the same direction as the peak of the VHE emission.
   Assuming that \mbox{MAGIC J1857.6+0297} is indeed associated with this cloud complex,
   the overall lack of overlap with the most dense parts of the molecular clouds hints at a leptonic origin for the VHE gamma-ray emission.
   A hadronic origin is not excluded, however, due to the possible presence of an atomic gas cloud within the proposed cavity.
   We have outlined alternative scenarios that could explain the proposed link between the gamma-ray emission and this molecular cloud complex.
   They include a PWN whose progenitor star created a cavity in the cloud prior to its SN explosion,
   particle acceleration based on the combined effects of strong stellar winds from massive stars,
   or the acceleration of particles in the outflows of massive protostars.
   More targeted multi-wavelength data will be needed, especially in X-rays,
   in order to characterize this interesting region and test whether it really is the source of the TeV gamma rays.
   
\begin{acknowledgements}
      We would like to thank the Instituto de Astrof\'{\i}sica de
      Canarias for the excellent working conditions at the
      Observatorio del Roque de los Muchachos in La Palma.
      The support of the German BMBF and MPG, the Italian INFN, 
      the Swiss National Fund SNF, and the Spanish MICINN is 
      gratefully acknowledged.
      This work was also supported by the CPAN CSD2007-00042 and MultiDark
      CSD2009-00064 projects of the Spanish Consolider-Ingenio 2010
      programme, by grant 127740 of the Academy of Finland,
      by the DFG Cluster of Excellence ``Origin and Structure of the 
      Universe'', by the Croatian Science Foundation Project 09/176,
      by the DFG Collaborative Research Centers SFB823/C4 and SFB876/C3,
      and by the Polish MNiSzW grant 745/N-HESS-MAGIC/2010/0.
\end{acknowledgements}


\begin{appendix}
  \section{Multi-wavelength data}

  Here we provide a technical description of the multi-wavelength data used in section ~\ref{sec:mwl}.
  The VGPS H{\sc i} data \citep{Stil2006} consists of continuum images with angular resolution of $1'$ and an \emph{rms} noise of $0.3$ K,
  while the H{\sc i} line images also have an angular resolution of $1'$,
  a spectral resolution of $1.56$ km s$^{-1}$ and an \emph{rms} noise of $0.3$ K per channel ($0.824$ km s$^{-1}$).
  For the  \mbox{$^{13}\textrm{CO}(\textrm{J}=1\rightarrow0)$} data,
  the  GRS survey \citet{Jackson2006} employed a grid spacing of $22''$, and had an angular resolution of $46''$
  and a spectral resolution of $0.212$ km s$^{-1}$, with \emph{rms} noise of $\thicksim 0.13$ K.
  Finally, the $8$ $\mu$m  GLIMPSE data \citep{Benjamin2013} has arcsecond-scale resolution.
  
  Our analysis of the radio data cubes used the \emph{Miriad} \citep{Sault1995} package,
  while the images in this section were generated using the \emph{Karma} \citep{Gooch1996} and \emph{DS9} \citep{joymandel2003} software packages.
  
  \section{Avergae \mbox{$^{13}\textrm{CO}(\textrm{J}=1\rightarrow0)$} emission}
  
\begin{figure}
    \includegraphics[height=.14\textheight]{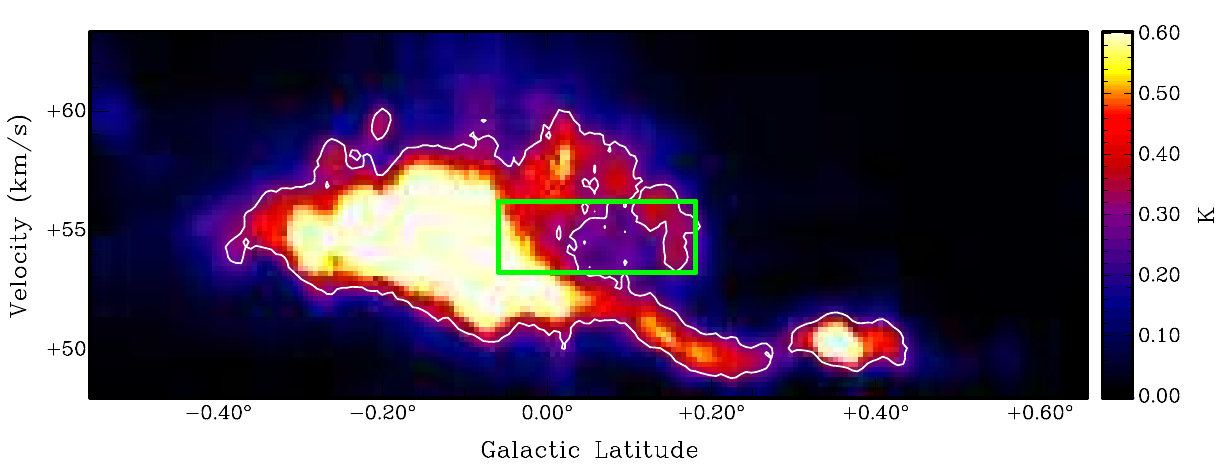}
    \includegraphics[height=.14\textheight]{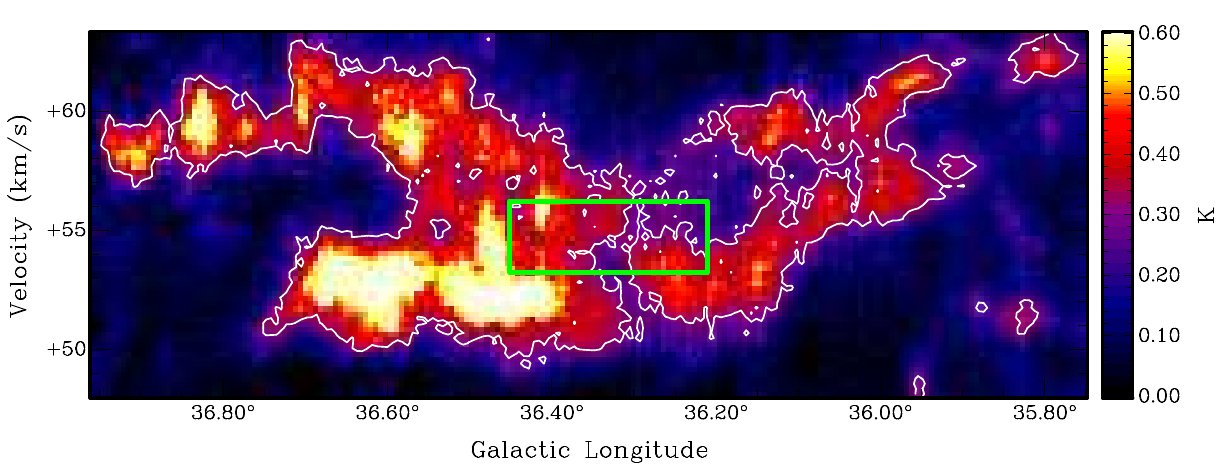}
    \caption{
      Galactic latitude-velocity (left) and longitude-velocity (right) plots
      showing \mbox{$^{13}\textrm{CO}(\textrm{J}=1\rightarrow0)$} line emission
      averaged over the range of longitudes: \mbox{$36.21^{\circ} - 36.45^{\circ}$ (left)}
      and latitudes: \mbox{$-0.06^{\circ} - +0.21^{\circ}$ (right)}.
      The green rectangles indicate the approximate location of a possible gas cavity.
      Their angular positions correspond to the extent of the dashed circle in \mbox{Figure \ref{figco}}
      while in velocity they span the $V_{LSR}$ range of $53.23 - 56.21$ km s$^{-1}$.
      The white contours indicate levels of $0.3$ K.
    }
    \label{figlbv}
  \end{figure}

  \mbox{Figure \ref{figlbv}} shows Galactic longitude-velocity and latitude-velocity plots of
  \mbox{$^{13}\textrm{CO}$} emission in the direction of a possible gas cavity that is discussed in Section 4.

\end{appendix}

\end{document}